\documentclass[11pt,preprint,preprintnumbers,a4paper]{article}
\pdfoutput=1

\usepackage{jheppub}
\usepackage{graphicx,array}
\usepackage{color}
\usepackage{latexsym}
\usepackage{amsthm}
\usepackage{amsmath}
\usepackage{amssymb}
\usepackage{hyperref}
\usepackage{bbold}
\usepackage{subfig}
\usepackage{multirow}
\usepackage{slashed}
\usepackage[normalem]{ulem}
\usepackage{hyperref}

\newcommand{\nc}{\newcommand}

\nc{\beq}{\begin{equation}}  
\nc{\eeq}{\end{equation}}  
\nc{\beqa}{\begin{eqnarray}}  
\nc{\eeqa}{\end{eqnarray}}  
\nc{\bit}{\begin{itemize}}  
\nc{\eit}{\end{itemize}}

\title{
Muon $g-2$ from Millicharged Hidden Confining Sector
}

\author[a]{Yang~Bai,}
\author[b]{Seung~J.~Lee,}
\author[c]{Minho~Son,} 
\author[c]{and Fang~Ye}

\affiliation[a]{Department of Physics, University of Wisconsin-Madison, Madison, WI 53706, USA}
\affiliation[b]{Department of Physics, Korea University, Seoul 136-713, Korea }
\affiliation[c]{Department of Physics, Korea Advanced Institute of Science and Technology, \\
291 Daehak-ro, Yuseong-gu, Daejeon 34141, Republic of Korea}

\emailAdd{yangbai@physics.wisc.edu}
\emailAdd{s.jj.lee@gmail.com}
\emailAdd{minho.son@kaist.ac.kr}
\emailAdd{shivaandalice@gmail.com}

\abstract{We provide a novel explanation to the muon $g-2$ excess with new physics contributions at the two-loop level.  In this scenario, light millicharged particles are introduced to modify the photon vacuum polarization that contributes to muon $g-2$ at one additional loop. The muon $g-2$ excess can be explained with the millicharged particle mass $m_\chi$ around 10 MeV and the product of the multiplicity factor and millicharge squared of $N_\chi \varepsilon^2 \sim 10^{-3}$. The minimal model faces severe constraints from direct searches at fixed-target experiments and astrophysical observables. However, if the millicharged particles are also charged under a hidden confining gauge group $SU(N_\chi)$ with a confinement scale of MeV, hidden-sector hadrons are unstable and can decay into neutrinos, which makes this scenario consistent with existing constraints. This explanation can be well tested at low-energy lepton colliders such as BESIII and Belle II as well as other proposed fixed-target experiments.
}

\begin{document}

\maketitle 

\section{Introduction}
\label{sec:intro}

As one of the most precisely measured quantities in particle physics, the muon anomalous magnetic moment provides a testing ground for the  Standard Model (SM) of particle physics.
The recent result from Fermi National Laboratory (FNAL) Muon $g-2$ collaboration confirms the previous measurement from Brookhaven National Laboratory (BNL) E821 experiment. The combined experimental measurements now have 4.2$\sigma$ tension with the SM prediction and have the difference of 
\begin{equation}\label{eq:mug2:2021:comb}
\Delta a_\mu^{\text{FNAL+BNL}} = a_\mu^{\text{exp}} - a_\mu^{\text{SM}}  = 251(59) \times 10^{-11}~,
\end{equation}
with $a_\mu \equiv (g-2)_\mu/2$. Here, the individual deviations from FNAL and BNL are $\Delta a_\mu^{\text{FNAL}} = 230(69)\times 10^{-11}$ and $\Delta a_\mu^{\text{BNL}} = 279(76)\times 10^{-11}$, respectively~\cite{Bennett:2006fi,Keshavarzi:2018mgv,Abi:2021gix}. 

This new result suggests possible new physics beyond the SM with new particles existing at the collider-accessible mass range. In the literature, new particles have been introduced to directly couple to  muon with the new physics contributions happening at one-loop level (a cancelation among one-loop contributions could lead to a dominant two-loop contribution~\cite{Balkin:2021rvh}). In this paper, we will explore an alternative explanation without a direct coupling of new particles to  muon and have the new physics contributions at the two-loop level. Before we mention our scenario, we briefly discuss some SM two-loop contributions to the muon $g-2$.

At the two-loop level, one important SM contribution is the so-called hadronic vacuum polarization (HVP) with hadrons contributing to photon vacuum polarization and then to $a_\mu$. There are two different ways to obtain this contribution: R-ratio and Lattice QCD. For the first R-ratio way, experimental data for the $e^+ e^- \rightarrow \gamma^* \rightarrow \mbox{hadrons}$ are collected and used to derive the photon vacuum polarization from a dispersion relation~\cite{Hagiwara:2003da,Davier:2017zfy,Colangelo:2018mtw,Hoferichter:2019mqg,Davier:2019can}. The updated results have $a_\mu^{\text{LO-HVP}}(\mbox{R-ratio})  = 6931(40)\times 10^{-11}$~\cite{Aoyama:2020ynm}, which provides the dominant contribution to the total error of the SM prediction $a_\mu^{\text{SM}} = 116591810(43) \times 10^{-11}$~\cite{Aoyama:2020ynm}. On the Lattice QCD side, the world averaged value has $a_\mu^{\text{LO-HVP}}(\mbox{Lattice})  = 7116(184)\times 10^{-11}$~\cite{Aoyama:2020ynm}, without including the BMW-2020 Lattice result with $7075(55)\times 10^{-11}$~\cite{Borsanyi:2020mff}, which suggests a smaller discrepancy between the SM prediction and the experimental result and is to be confirmed by other Lattice simulations. 

It is interesting to notice that the reported deviation in Eq.~\eqref{eq:mug2:2021:comb} corresponds to roughly 4\% of the aforementioned HVP contribution in the SM, $\Delta a_\mu^{\text{FNAL+BNL}}  \sim 4\% \times a_\mu^{\text{LO-HVP}}(\text{SM})$. 
If new physics modifies the photon vacuum polarization and then $\Delta a_\mu$, its contribution could be at the percent level compared to the SM contribution. One obvious requirement for this type of new physics is that they must couple to the photon or have electromagnetically-charged particles. On the other hand, the new particle does not necessarily behave as hadrons or contribute to R-ratio. For instance, they could have a milli-electric charge and behave as a missing particle at colliders (we do not consider another interesting milli-magnetically-charged particle case here). 

\begin{figure}[t!]
\begin{center}
\includegraphics[width=0.45 \textwidth]{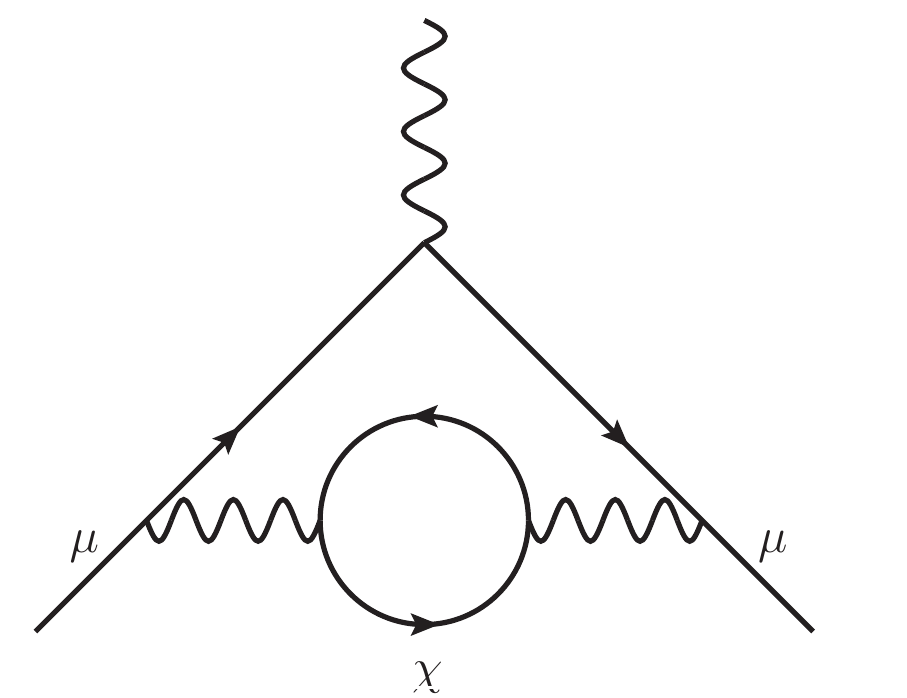}
\caption{The two-loop contribution to muon $g-2$ from millicharged particles $\chi$ contributions to the photon vacuum polarization. 
\label{fig:vap:mCP}}
\end{center}
\end{figure}

More specifically, as an alternative explanation of the muon $g-2$ excess, we introduce millicharged particles (mCP's) with a mass $m_\chi$ and a multiplicity factor $N_\chi$. We first assume that the millicharged particles have only electromagnetic interactions and will introduce additional hidden confining gauge interactions for them later in order to be consistent with various constraints.
Their interaction with the photon has the form of $ \varepsilon\,e \,A_\mu \overline\chi \gamma^\mu \chi$, with $\varepsilon < 1$ as the electric charge of the new $\chi$ particles. For simplicity, we choose the same charge and mass for all millicharged particles. 
We also keep the origin of the small charge $\varepsilon$ agnostic (note that several possiblilities has been considered in the literature, including the millicharged particles in grand unified theory~\cite{Okun:1983vw}, models with a paraphoton~\cite{Holdom:1985ag} and milli-electrically-charged neutrinos~\cite{Foot:1992ui}) and explore mainly the phenomenological consequence here. The dominant two-loop contribution to the muon $g-2$ from $\chi$ is illustrated in Fig.~\ref{fig:vap:mCP}. Because the lepton mass flips the helicity, the corresponding contribution to the electron $g-2$ is suppressed by a factor of $(m_e/m_\mu)^2$ for the mass range of interest and can be consistent with the measured value. 

%

Our paper is organized as follows. In Section~\ref{sec:muong2}, we estimate the two-loop QED correction to the muon (electron) $g-2$ from the millicharged particles, and compare them with the observed muon (electron) $g-2$ result. In Section~\ref{sec:detection}, we derive various parametric dependences of the signal rates on $m_\chi$, $\varepsilon$, and $N_\chi$ and use them to recast results from various low-energy collider, fixed-target and neutrino experiments. We also discuss the astrophysical constraints. In Section~\ref{sec:hidden-confine}, the hidden confining gauge interactions are introduced for $\chi$ and various constraints are revisited. We summarize our results in Section~\ref{sec:conclusions}.  In Appendix~\ref{app:sec:alphaMZ}, we discuss the modification of the running fine structure constant due to millicharged particles. In Appendix~\ref{app:sec:twoloopQED}, we quote the formula of the two-loop QED correction from a lepton to muon $g-2$. In Appendix~\ref{app:sec:darkQCD}, we present some formulas for the properties of hidden hadrons. 

\section{Muon $g-2$ from mCP}
\label{sec:muong2}

The contribution to the muon $g-2$ through the one-loop vacuum polarization of mCPs  is analogous to the QED contribution from SM leptons to the muon $g-2$ up to an overall factor of $N_\chi \varepsilon^2$. Due to the coupling to photon, the experimental limits on the millicharged particles are stringent. While we postpone the detailed experimental constraints to the next section, here we simply report the constrained upper limit as $N_\chi \varepsilon^2 \lesssim 6.4\times 10^{-3}$ for $m_\chi \sim 15$~MeV, which will guide us for discussing the muon $g-2$. 

\begin{figure}
\begin{center}
\includegraphics[width=0.55 \textwidth]{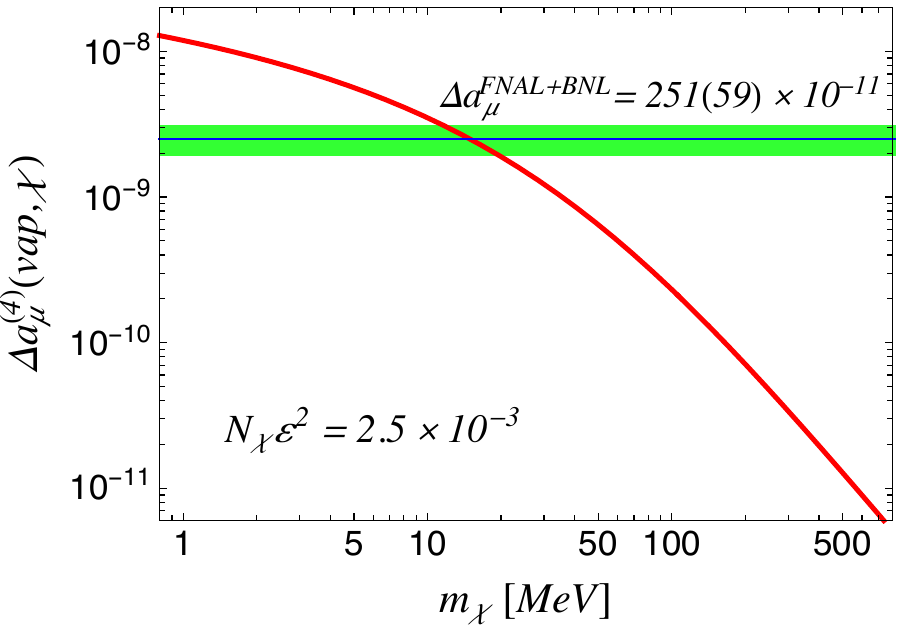}
\caption{The two-loop contribution to muon $g-2$ from millicharged particles through one-loop photon vacuum polarization for $N_\chi \varepsilon^2 = 2.5\times 10^{-3}$. The green band corresponds to $(251\pm 59) \times 10^{-11}$.
\label{fig:vap:mCP:numeric}}
\end{center}
\end{figure}

Using the exact expression~\cite{Elend:1966,Passera:2004bj,Passera:2006gc} (see Appendix~\ref{app:sec:twoloopQED}), the numerical value of $\Delta a_\mu$ as a function of mCP mass $m_\chi$ is shown in Fig.~\ref{fig:vap:mCP:numeric} for $N_\chi \varepsilon^2=2.5\times 10^{-3}$.  Apparently, millicharged particles with $m_\chi \sim 15$ MeV  and $N_\chi \varepsilon^2  \simeq 2.5\times 10^{-3}$ can account for the observed $\Delta a_\mu$ through the photon vacuum polarization.
For $m_\chi < m_\mu$, one can approximate $\Delta a_\mu$ as 
\begin{eqnarray}
\begin{split}
  \Delta a_\mu^{(4)} (\text{vap},\chi)   &\approx \ N_\chi \varepsilon^2 \left  [ -\frac{1}{3} \log \frac{m_\chi}{m_\mu} - \frac{25}{36}  
   + \frac{\pi^2}{4} \frac{m_\chi}{m_\mu} +\left ( 3 + 4 \log \frac{m_\chi}{m_\mu} \right ) \left ( \frac{m_\chi}{m_\mu} \right )^2 \right ] \left (\frac{\alpha}{\pi} \right )^2~
  \\[2pt]
  &\rightarrow\ 251 \times 10^{-11} \cdot \Big ( \frac{N_\chi \varepsilon^2}{2.52 \times 10^{-3}} \Big )\qquad (m_\chi = 15\ \text{MeV})~,
\end{split}
\end{eqnarray}
where superscript 4 (vap) denotes the order of the QED interaction (vacuum polarization).

While the millicharged particles can also contribute to the electron $g-2$ with the same overall sign,  the experimental result is smaller than the SM prediction~\cite{Parker:2018vye}
\begin{equation}
 \Delta a_e =  a_e^{\text{exp}} - a_e^{\text{SM}} = - 88(36) \times 10^{-14}~.
\end{equation}
For a heavier $\chi$ mass, the contribution to the electron $g-2$ is easily suppressed. Using the heavy mass approximation for $m_\chi \gtrsim m_\mu \gg m_e$,
\begin{equation}\label{eq:muong2:chi:heavy}
\begin{split}
  \Delta a_\mu^{(4)}(\text{vap},\chi) &\approx  \frac{N_\chi \varepsilon^2}{45} \Big ( \frac{m_\mu}{m_\chi} \Big )^2 \Big ( \frac{\alpha}{\pi} \Big )^2
\quad \,  (m_\chi \gtrsim m_\mu)~.
\end{split}
\end{equation}
The muon and electron $g-2$ are related by
\begin{equation}
\begin{split}
  \Delta a_e^{(4)}(\text{vap},\chi)
     &\approx  \left ( \frac{m_e}{m_\mu} \right )^2 \Delta a_\mu^{(4)}(\text{vap},\chi)\ \  \  \quad (m_\chi > m_\mu)~,       
\end{split}
\end{equation}
where $(m_e/m_\mu)^2 = 2.34 \times 10^{-5}$. For $m_\mu > m_\chi \gg m_e$ and as evident from Fig.~\ref{fig:vap:mCP:numeric}, the scaling behavior of $\Delta a_\mu^{(4)}(\text{vap},\chi)$ in $m_\chi$ deviates from the heavy $m_\chi$ approximation in Eq.~\eqref{eq:muong2:chi:heavy}. Numerically, the electron $g-2$ measurement sets a lower bound on $m_\chi$, or
\begin{equation}
\begin{split}
  \Delta a_e^{(4)}(\text{vap},\chi) 
  &\approx 2.8045 \times 10^{-12} \cdot  N_\chi \varepsilon^2  \cdot \Big (\frac{m_\mu}{m_\chi} \Big )^2
  \lesssim  | a_e^{\text{exp}} - a_e^{\text{SM}}| = 88 \times 10^{-14}~,
\end{split}
\end{equation}
where the numerical value in the first line corresponds to the $\chi$ contribution to electron $g-2$ in terms of the mass ratio $m_\mu$ to $m_\chi$. This formula implies that the case with $m_\chi \ll \mathcal{O}(m_\mu)$ requires a much smaller value of $N_\chi \varepsilon^2$, which may not be large enough to explain $\Delta a_\mu$. As can be seen from Fig.~\ref{fig:vap:mCP:numeric:2D}, the electron $g-2$ measurement in conjunction with the $\Delta a_\mu$ excess requires $m_\chi > 7$~MeV. 

There are additional higher-loop and suppressed contributions to $\Delta a_\mu$ that millicharged particles can  provide. The three-loop and mixed contribution from both $\chi$ and the SM leptons $e,\mu,\tau$ to photon vacuum polarization occurs at the $(\alpha/\pi)^3$ order and hence is small. The three-loop light-by-light contribution from an mCP loop is suppressed by an extra $\varepsilon^2 \alpha/\pi$ compared to the HVP contribution and can also be ignored.

One might expect that copious millicharged particles contributing to the photon vacuum polarization may also be constrained by the electroweak (EW) precision observables from the modified value of the fine structure constant at the electroweak scale. We have found that a light $m_\chi \ll m_{\pi^\pm}$ can explain the muon $g-2$ excess and in the meanwhile be consistent with the inferred value or precision of $\alpha(M_Z^2)$ from EW precision data. This is due to the power-law enhanced contribution to $a_\mu$ for a light $m_\chi$ as can be seen from Fig.~\ref{fig:vap:mCP:numeric}, while $\alpha(M_Z^2)$  only logarithmically depend on $m_\chi$. The detailed discussion on $\alpha(M_Z^2)$ due to mCPs is provided in Appendix~\ref{app:sec:alphaMZ}. In Fig.~\ref{fig:vap:mCP:numeric:2D} we show the black dashed line for the mCP contribution to $\alpha(M_Z^2)$ at the fraction of $0.01\%$, which is approximately the precision on the extracted $\Delta \alpha^{(5)}_{\text{had}} (M_Z^2)$ from EW precision data~\cite{Crivellin:2020zul}.

\section{Constraints on MCP}
\label{sec:detection}

For millicharged particles with a mass below $\sim 1$~MeV, the Big Bang nucleosynthesis (BBN) physics provides very stringent constraints on the charge~\cite{Davidson:1993sj,Davidson:2000hf}. Therefore, we will focus on the mCP masses above 1 MeV in this paper. For this range of masses, mCPs have been searched for at both collider and fixed-target experiments. One could group the searches into two categories: I) {\it the indirect way} without detecting the interaction-generated signals when the produced mCPs propagate in matter. II) {\it the direct way} to detect the mCP-electron or nucleon scattering generated signals. 

For the indirect way, we first note that searches for invisible decays of hadron with J$^{\rm PC}=1^{- -}$ can set an upper bound on $N_\chi \,\varepsilon^2$, which is the same parameter dependence as the one for $\Delta a_\mu$. Two states, $J/\psi$ and $\Upsilon(1S)$, stand out as the two most relevant ones. For $\Upsilon(1S)$, one has $\mbox{Br}\left(\Upsilon(1S) \rightarrow e^+ e^-\right) \approx (2.38\pm 0.11)\%$ and $\mbox{Br}\left(\Upsilon(1S) \rightarrow \mbox{invisible} \right) < 3.0\times 10^{-4}$~\cite{Aubert:2009ae}. The ratio of those two provide us a constraint on the mCP model parameter 
\beqa
N_\chi \varepsilon^2  = \frac{\mbox{Br}\left(\Upsilon(1S) \rightarrow \mbox{invisible} \right) }{\mbox{Br}\left(\Upsilon(1S) \rightarrow e^+ e^-\right)} < 1.3 \times 10^{-2} ~,
\eeqa
for $m_\chi \ll M_{\Upsilon(1S)}/2 \approx 4.7$~GeV to ignore the phase space factor. The upper limit from $J/\psi$  invisible decay is very similar to the above one. 

\begin{figure}
\begin{center}
\includegraphics[width=0.50 \textwidth]{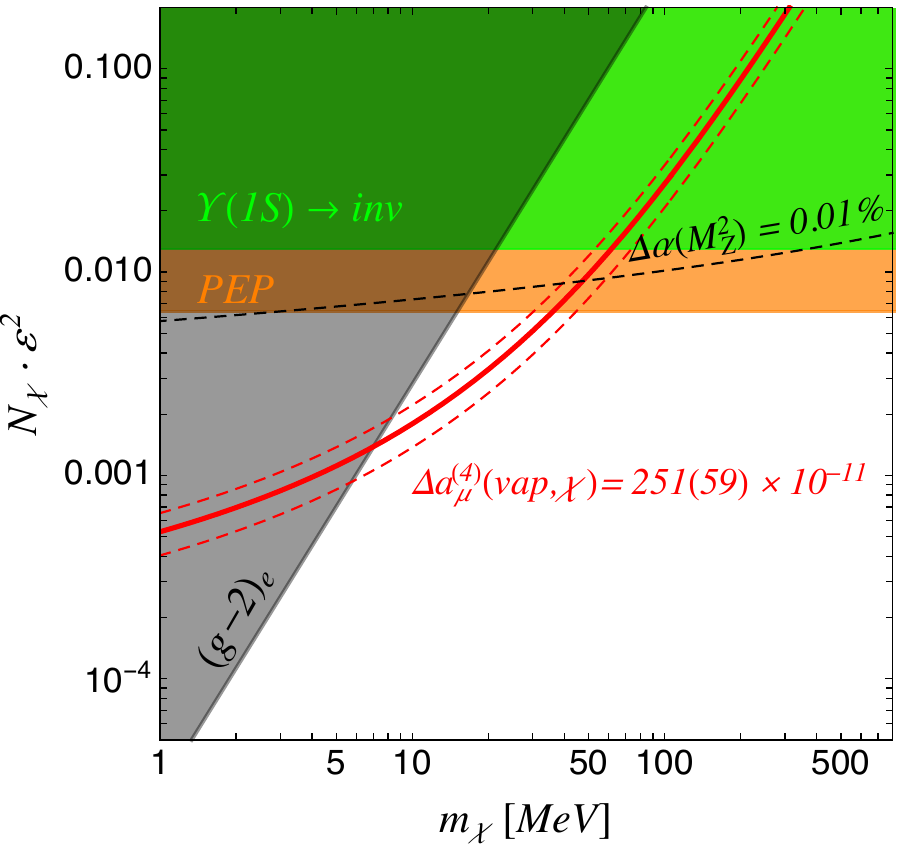} 
\caption{
The two-loop contribution to muon $g-2$ from millicharged particles from their one-loop contributions to photon vacuum polarization (red solid lines) and various experimental constraints that only depend on $N_\chi \varepsilon^2$. Below the black dashed line, the modification to the fine structure constant at the $Z$ boson mass, $\alpha(M_Z)$, is smaller than 0.01\%. 
\label{fig:vap:mCP:numeric:2D}}
\end{center}
\end{figure}

Another indirect  search for mCPs is to use signatures of mono-photon plus missing energy or $e^+e^- \rightarrow \gamma + \slashed{E}$ at a lepton collider. Historically, the PEP (Positron-Electron Project) experiment at SLAC has searched for $\gamma\, +$ weakly interacting particles with the constraint~\cite{Davidson:1991si}
\beqa
N_\chi \varepsilon^2 < 6.4 \times 10^{-3} ~, \qquad (\mbox{PEP}) 
\eeqa
for $m_\chi < 5$~GeV. 
Other low-energy lepton colliders such as BaBar~\cite{Aubert:2008as}, BESIII~\cite{Liu:2018jdi} and Belle II  could have impressive projected limits of $N_\chi \varepsilon^2  \lesssim 10^{-6}$ for $m_\chi < 1$~GeV if there is no additional larger backgrounds as considered in Ref.~\cite{Liang:2019zkb}. Additionally, searches for events with large missing energy in the fixed-target experiment with an electron beam can also be used to constrain the millicharged particles, if one recasts the limits on produced on-shell invisible dark photon into the signals with two mCPs in the NA64 experiment~\cite{NA64:2016oww,Banerjee:2019pds}. 
Dedicated experimental searches are required to cover the $\Delta a_\mu$-preferred parameter space. 
We summarize various indirect constraints in Fig.~\ref{fig:vap:mCP:numeric:2D}. One can see from this plot that the $\Delta a_\mu$-preferred millicharged particle mass is constrained to be $7\,\mbox{MeV} \lesssim m_\chi \lesssim 36\,\mbox{MeV}$.

For the direct way of searching for mCPs at a fixed-target experiment, the signal events have a simple scaling in terms of $N_\chi$ and $\varepsilon$ as 
\beqa
N_{\rm signal} \propto N_\chi \, \varepsilon^{2 + 2 n_{\rm hit}} ~, 
\eeqa
with $n_{\rm hit}$ as the number of hits required for a certain experiment to reduce backgrounds. The factor of $N_\chi \varepsilon^2$ comes from the productions of  mCPs either from meson decays or Drell-Yan processes. The remaining factor $\varepsilon^{2 n_{\rm hit}}$, with usually $n_{\rm hit} = 1, 2$, comes from the detecting probability. For any experimental constraint $\varepsilon < \varepsilon_{\rm max}(N_\chi = 1, m_\chi)$ for the case with $N_\chi = 1$, we can reinterpret it for the large $N_\chi$ case as 
\beqa
\label{eq:reinterpret} 
N_\chi \varepsilon^2 < N_\chi^{\frac{n_{\rm hit}}{1+n_{\rm hit}} }\, \varepsilon^2_{\rm max}(N_\chi =1, m_\chi) ~. 
\eeqa

For $m_\chi  < 100$~MeV, the most stringent constraint comes from the MilliQ experiment~\cite{Prinz:1998ua} at SLAC. For $N_\chi = 1$, the constraint is approximately $\varepsilon < (1.7\times 10^{-4}) \times (m_\chi / 15\,\mbox{MeV})^{1/2}$ for $m_\chi  < 100$~MeV. Since a singlet hit has been required, one can reinterpret the constraint to copious mCPs as 
\beqa
N_\chi\, \varepsilon^2 &<& N_\chi^{1/2}\times  (1.7\times 10^{-4})^2 \, \times \left( \frac{m_\chi}{15\,\mbox{MeV}}\right)\,, 
 \hspace{0.1cm} \mbox{for} \ \ m_\chi < 100\,\mbox{MeV}  \quad \mbox{(MilliQ@SLAC)} ~.  \nonumber
\eeqa
For $m_\chi = 15$~MeV, the $\Delta a_\mu$-preferred $N_\chi \varepsilon^2 \approx 2.5\times 10^{-3}$ requires a lower bound on the multiplicity $N_\chi > 7.4\times 10^{9}$. 

For a heavier mass above 100 MeV, both MiniBooNE~\cite{Magill:2018tbb} with a single hit and ArgoNeuT~\cite{Harnik:2019zee,Acciarri:2019jly} with doublet hits provide the most stringent constraints for the $N_\chi =1$ case. For the MiniBooNE, the reinterpretation of the limits for the $N_\chi =1$ is similar to the MilliQ experiment. For the constraints from ArgoNeuT with $n_{\rm hit} = 2$, the constraint is $\varepsilon_{\rm max}(N_\chi =1, m_\chi) \approx 4.5\times 10^{-3}$ for $100\,\mbox{MeV} \le m_\chi \le 200\,\mbox{MeV}$. So, for $m_\chi = 200$\,MeV and  to explain $\Delta a_\mu$, one requires $N_\chi > 3\times 10^5$, although this mass range has already been excluded by the indirect way. A similar mass range can also be probed by an atmospheric millicharged particle search~\cite{Plestid:2020kdm,ArguellesDelgado:2021lek}.

Given the large multiplicity factor $N_\chi \gtrsim 10^{10}$, one should check the consistency of our explanation for the muon $g-2$ excess against cosmological~\cite{Dubovsky:2003yn} and astrophysical bounds~\cite{Davidson:1993sj,Davidson:2000hf,Chang:2018rso}. The mCPs with mass $m_\chi\sim\mathcal O(10\,\mbox{MeV})$ could be constrained from the supernova (SN) 1987A observations~\cite{Davidson:1993sj,Davidson:2000hf,Chang:2018rso}. If mCPs interact negligibly with the SN plasmon once they are produced, they will free stream, leave the SN, take energy away and serve as an additional cooling channel for the SN. If mCPs interact with the plasmon sufficiently, they will become trapped and only escape via their last scattering surface given by the trapping radius, with a much lower temperature than the core. A conservative requirement is to have the energy carried away by mCPs to be smaller than that from neutrinos. The latest bounds from the above two cases are updated in Ref.~\cite{Chang:2018rso}, which can be applied to the mCP model with a large multiplicity. For the free-streaming case, the luminosity of the mCPs $\propto N_\chi \varepsilon^2$ and provides a constraint $N_\chi \varepsilon^2 \lesssim 10^{-16}$  for $m_\chi \sim 15~\mbox{MeV}$, which excludes the muon $g-2$ preferred region. In the trapped case, the luminosity of the mCPs $\propto N_\chi \varepsilon^{-\beta}$, with $11<\beta<14$, should be smaller than around $(7\times 10^{-6})^{-\beta}$.  The inequality can be converted to $\varepsilon \gtrsim 7\times 10^{-6}\times N_\chi^{1/\beta}$, which give $N_\chi \,\varepsilon^2 > (7\times 10^{-6})^2\times N_\chi^{1+2/\beta}$. Combining with the other bound $N_\chi > 7.4\times 10^9$, this excludes the muon $g-2 $ preferred parameter region $N_\chi \varepsilon^2\sim 10^{-3}$.

\section{Millicharged Hidden Confining Sector}
\label{sec:hidden-confine}

Given the stringent constraints for the minimal millicharged particle model, we introduce a hidden confining gauge group
$SU(N_\chi)$ with $\chi_{L, R}$ charged as the fundamental representation of $SU(N_\chi)$ (see Ref.~\cite{Kribs:2016cew} for a review). The confinement scale $\Lambda_\chi$ of $SU(N_\chi)$ is assumed to be lower than the bare quark mass, or $\Lambda_\chi \ll m_\chi$.  We will choose $\Lambda_\chi = \mathcal{O}(\mbox{1\,MeV})$ to avoid any possible constraints from BBN physics at the temperature of around 1 MeV. So, the gauge-singlet boundstates contain various ``quarkonium"-like $\chi\overline{\chi}$ meson states, glueballs and millicharged baryonic states. To satisfy the SN cooling bound as well as direct search constraints for the meson states in fixed-target experiments, we introduce a scalar particle $S$ to mediate the decays of glueball states into neutrinos. The relevant Lagrangian terms are
\beqa
\label{eq:Lag-confining}
\mathcal{L} &\supset& - \frac{1}{4} G_{
\chi\,a\,\mu\nu}G_\chi^{a\,\mu\nu} +  \overline{\chi} i \gamma_\mu \left(\partial^\mu - i \varepsilon A^\mu - i g_\chi \,t^a\,G^{a\,\mu}_\chi\right) \chi  \nonumber \\
&& - \, m_\chi\,\overline{\chi} \chi\, - \frac{1}{2}m_S^2 S^2 \, - \, i\,y_s\,S\,\overline{\chi}\gamma_5 \chi \, 
  - \frac{S(LH)^2}{\Lambda^2} ~.
\eeqa
Here, $G_\chi^{a\,\mu}$ is the hidden gauge field; $t^a$ is the group generator; $y_s$ is the Yukawa coupling for the light scalar $S$ coupling to $\chi$; the last term is the dimension-6 operator for $S$ coupling to neutrinos after electroweak symmetry breaking; the scalar mass is taken to be smaller than the confinement scale, or $m_S < \Lambda_\chi \ll m_\chi$, such that the glueball states with a mass $\sim \Lambda_\chi$ can decay into two $S$'s which eventually decay into neutrinos. 
The cutoff scale $\Lambda$ in Eq.~\eqref{eq:Lag-confining} is parametrically lower than the typical lepton-number violating scale in the SM associated with the dimension-5 neutrino mass operator, $(LH)^2$. The $(LH)^2$ operator is suppressed by an approximate discrete $\mathcal{Z}_2$ symmetry under $S \rightarrow -S$, $\chi_{L/R} \rightarrow \chi_{R/L}$, $L \rightarrow iL$, $e_R \rightarrow i e_R$, and $H\rightarrow H$.
The calculated decay widths for the unstable meson and glueball states can be found in Appendix~\ref{app:sec:darkQCD}. For our purpose here, the $J^{PC}=1^{--}$ $\chi\overline{\chi}$ boundstate, $\Upsilon_\chi$ with a mass $m_{\Upsilon_\chi} \approx 2 m_\chi$, mainly decays into glueballs, $\mathcal{G}_\chi$, which has the dominant decay channel into two $S$'s that eventually decay into neutrinos.

For this hidden-confining millicharged model, the indirect constraints in Section~\ref{sec:detection} still apply and the allowed parameter space is identical to Fig.~\ref{fig:vap:mCP:numeric:2D}. On the other hand, the direct detection constraints do not apply here, because the produced $\chi$ and $\overline{\chi}$ hadronize first into mesons that eventually decay into neutrinos and evade the detections. For the sub-leading channels into electrons or photons, the material between the production and detector regions absorbs them. For the SN cooling constraints, we first note that the stable and millicharged baryonic state $\mathcal{B}_\chi$ with a charge $N_\chi \, \varepsilon$ and a mass $\approx N_\chi\,m_\chi$ can escape the plasma. However, for a large $N_\chi \gtrsim 10$, the baryon mass could be much larger than the SN core temperature $T_{\rm SN} \sim 30$~MeV, so its production is exponentially suppressed by a factor of $e^{- m_{\mathcal{B}_\chi}/T_{\rm SN}}$ and can be ignored. For the unstable meson and glueball states, we require their decay lengths to be shorter than the SN core radius of $\sim 1$~km. 
For the quarkonium states, this is easy to be satisfied because they decay almost instantaneously into glueball states. Using the expression for the decay width in Eq.~\eqref{eq:upsilon:to:hgluons}, its lifetime is estimated to be $c\tau_0(\Upsilon_\chi) \simeq 1.8 \times 10^{-8}~\mbox{cm}$ for a selected benchmark point, $m_\chi = 15$ MeV, $N_\chi = 10$, and $\Lambda_\chi = 3$ MeV.
For the glueball states, the $\mathcal{G}_\chi^{0^{++}}$ can promptly decay into two $S$'s via the interactions in Eq.~\eqref{eq:Lag-confining}, while for other states additional interactions may be needed to make them decay promptly.
For instance, the prompt decay of $\mathcal{G}_\chi^{0^{-+}}$ into two $S$'s is viable, provided the small $\mathcal{Z}_2$-symmetry-breaking Yukawa interaction $-\kappa_s y_s S \bar{\chi}\chi$ with $\kappa_s \ll 1$. 
Then, a $\mathcal{G}_\chi^{0^{-+}}$ hidden glueball can dominantly decay to two $S$'s via the dimension-6 operator, 
\begin{equation}
 N_\chi \frac{g_\chi^2 \kappa_sy_s^2}{16\pi^2 m_\chi^2} S^2 G_{\chi\, a\,\mu\nu} \widetilde{G}_\chi^{a\,\mu\nu}~,
\end{equation}
from which the lifetime for our benchmark point is estimated to be [see Eq.~\eqref{eq:0-+lifetime}]
\begin{equation}
\begin{split}
 c\tau_0(\mathcal{G}_\chi^{0^{-+}}) 
 \simeq 0.9\,\mbox{cm}\times\,\left ( \frac{10}{N_\chi} \right )^2 \left ( \frac{10^{-3}}{\kappa_s} \right )^2\,  \left ( \frac{2}{y_s} \right )^4\,\left ( \frac{m_\chi}{15\, \text{MeV}}
\right )^4\, \left ( \frac{3\,\text{MeV}}{\Lambda_\chi}
\right )^5~.
\end{split}
\end{equation}
For the $S$ state, it can decay into two neutrinos within the SN core radius $\sim 1$~km for a cutoff scale $\Lambda \lesssim 10^5$~GeV [see Eq.~\eqref{eq:S-lifetime}], 
\begin{equation}
\begin{split}
 &c\tau_0(S) = 
0.9 \,\mbox{cm} 
 \times \left ( \frac{1\, \text{MeV}}{m_S} \right ) \left ( \frac{\Lambda}{3\times 10^4\, \text{GeV}}
\right )^4~.
\end{split}
\end{equation}
With the above values of $\Lambda$ and $m_S$, $S$ decays into neutrinos well before the BBN time.
More detailed discussion can be found in Appendix~\ref{app:sec:darkQCD}.
\begin{figure}
\begin{center}
\includegraphics[width=0.50 \textwidth]{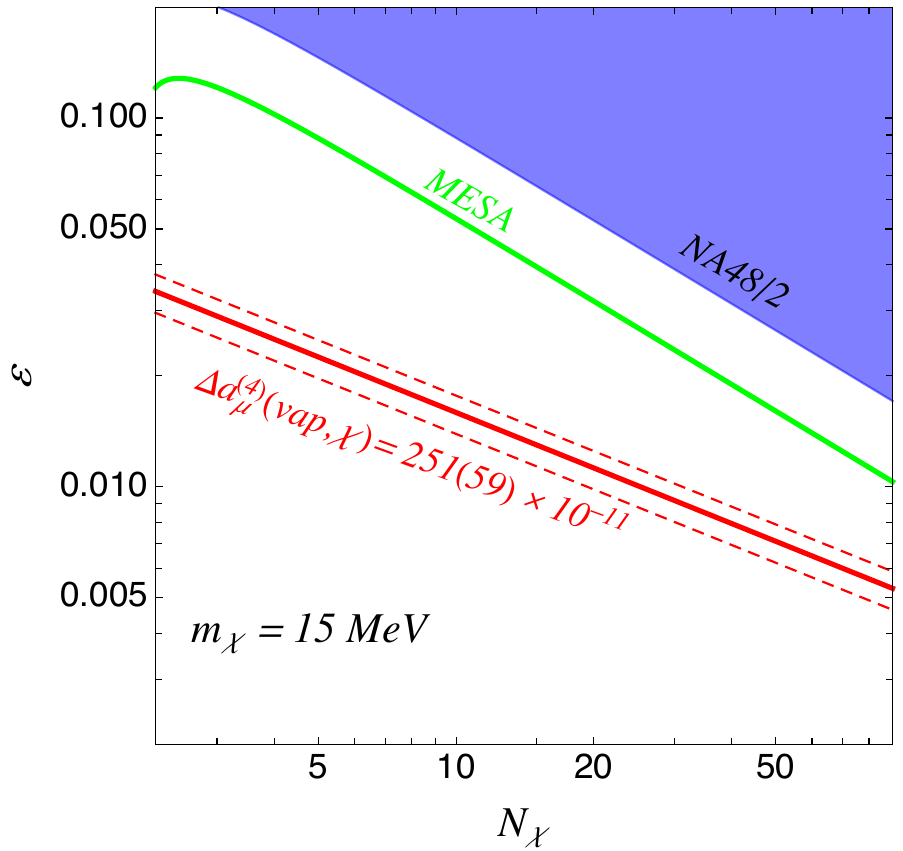} 
\caption{The parameter space in $\varepsilon$ and $N_\chi$ for the millicharged confining $SU(N_\chi)$ model. The light blue region is excluded by the NA48/2 experiment~\cite{Batley:2015lha}, while the green line is the projected limit from MESA~\cite{Doria:2019sux}. \label{fig:constrains:confining}}
\end{center}
\end{figure}

Although the hidden confining millicharged model can satisfy the previously mentioned constraints, there are additional constraints from fixed-target experiments that search for dark photon~\cite{Fabbrichesi:2020wbt}. For the model at hand, the $1^{--}$ $\Upsilon_\chi$ state has the same quantum number as photon and has a kinetic mixing term that is generated by $\chi$-loop diagrams. The effective kinetic parameter $-\frac{1}{2}\epsilon_{\rm kin} F_{\mu\nu}F^{\prime \mu\nu}$ has been derived in Appendix~\ref{app:sec:darkQCD}, and it is given by
\beqa
\epsilon_{\rm kin} = 2\,e\,\varepsilon\,\sqrt{N_\chi} \left (\frac{N_\chi^2-1}{2N_\chi} \frac{\alpha_\chi}{4} \right )^{3/2} ~,
\eeqa
with the gauge coupling $\alpha_\chi$ evaluated at the Bohr momentum of the boundstate.

For the beam-dump experiments like Orsay~\cite{Andreas:2012mt}  or NA64~\cite{Banerjee:2019hmi} that cover the dark photon mass range around $30$~MeV, the shielding length is 1 meter or above. The mainly produced hidden particles are the $\Upsilon_\chi$ state and/or its excited states. The productions for hidden baryonic states are negligible because they are suppressed by a higher power of  gauge couplings and a phase space factor. For the new particles from cascade decays of $\Upsilon_\chi$, their decay lengths could be much shorter than 1 meter even after including the Lorentz boost factor [see Eqs.~\eqref{eq:S-lifetime} and \eqref{eq:0-+lifetime}], which makes our model not constrained by those experiments. On the other hand, existing and future experiments like NA48/2~\cite{Batley:2015lha}, MESA~\cite{Doria:2019sux} and HPS~\cite{HPS:2018xkw}, do potentially cover the prompt decay scenario in our model. After taking into account of the production $\sigma_{\rm prod}(\Upsilon_\chi) \propto \epsilon_{\rm kin}^2$ and the branching ratio $\mbox{BR}(\Upsilon_\chi \rightarrow e^-e^+)$ [see Eq.~\eqref{eq:upsilon-branching-ee}], we define the following observable
\beqa
\begin{split}
\epsilon_{\gamma'}^2 &= \epsilon_{\rm kin}^2\, \text{BR} (\Upsilon_\chi \rightarrow e^+e^-) \
=\frac{3\pi^2\,\alpha^3\,\varepsilon^4\,N_\chi\,(N_\chi^2 - 1)^2}{8(\pi^2 - 9)\,(N_\chi^2 - 4)} ~, 
\end{split}
\eeqa
to approximately match the constraints on the square of the dark photon kinetic mixing parameter. We show the existing constraints and future limits in Fig.~\ref{fig:constrains:confining}.  One can see that the muon $g-2$ anomaly-preferred region for this $SU(N_\chi)$ confining model is still allowed by experimental constraints. Future experimental results from HPS~\cite{HPS:2018xkw} or others could cover the muon $g-2$ preferred parameter region.

Let us briefly comment on the relic abundance of the hidden baryon $\mathcal{B}_\chi$ (see also Ref.~\cite{Kribs:2009fy}). Since it has milli-electrical charge $N_\chi \,\varepsilon$ and interacts with other electrically charged visible particles, so it can only take a tiny fraction of cold dark matter (CDM). The current strongest upper bound on the fraction $f_{\text{mCP}} \equiv \rho_{\text{mCP}}/\rho_{\text{CDM}}$ arises from the CMB constraints, $f_{\text{mCP}} < 0.004$~\cite{Kovetz:2018zan}. As the temperature drops below the confinement scale, heavy (compared to the confinement scale), long-lived hidden baryons form. Hidden baryon and anti-baryon annihilation can be viewed as a $\mathcal{B}_\chi-\overline{\mathcal{B}}_\chi$ pair first forming a highly excited intermediate bound state and then this bound state losing energy and angular momentum via releasing $S$ particles and/or glueballs. The resulting annihilation cross section is geometric~\cite{Kang:2006yd} with $\langle \sigma v\rangle_{\text{ann}}\sim \pi R_{\text{had}}^2 (T_B / m_{\mathcal{B}_\chi})^{1/2}$, where $R_{\text{had}}$ is the hadron size and $T_B$ is the temperature at which the baryon is formed. Taking $R_\text{had}^{-1} \sim \Lambda_\chi\sim m_S\sim m_{\text{glueball}}$, we estimate the total fraction of hidden baryons as $f_{\rm mCP}\sim 7.2\times 10^{-13} \times (\Lambda_\chi/3\text{ MeV})^2 (m_\chi/15\text{ MeV})^{1/2} (3 \text{ MeV}/T_B)^{3/2}$, which easily satisfies the current constraint and only contributes a tiny fraction of dark matter.  

We also note that the phase transition for this hidden confining model with $\Lambda_\chi < m_\chi$ is a first-order one~\cite{Saito:2011fs}. The bubbles produced during the phase transition can also change the cosmological evolution of the hidden baryons~\cite{Asadi:2021pwo}. The stochastic gravitational waves produced during the phase transition with $T \sim \Lambda_\chi \sim$ few MeV could potentially explain the recent observations by the NANOGrav telescope~\cite{Arzoumanian:2021teu}.

\section{Conclusions}
\label{sec:conclusions}

We have provided a new scenario to explain the muon $g-2$ excess based on millicharged particle contributions to the photon vacuum polarization.  Similar to the two-loop contribution to the muon $g-2$ from the hadronic vacuum polarization in the SM, our model provides a hidden hadronic vacuum polarization contribution to the muon $g-2$. We have shown that lighter millicharged particles than the QCD pions can explain the muon $g-2$ excess, while being consistent with the inferred value or precision of the fine structure constant at the EW scale, or $\alpha(M_Z^2)$, from EW precision data, due to the power-law enhanced contribution to $a_\mu$ for a light millicharged particle [whereas $\alpha(M_Z^2)$ only logarithmically depend on the mass of the millicharged particle].

The simple millicharged particle model for the muon $g-2$ is excluded mainly due to the stringent cosmological bound from SN1987A. To avoid the SN cooling bound, the millicharged particle in this work is also charged under a hidden confining $SU(N_\chi)$ gauge group with its mass above the confinement scale. Millicharged particles form quarkonium-like bound states which instantaneously decay to hidden hadronic states (such as various hidden glueballs) with the largest branching fraction.
The hidden hadrons in this model mainly decay into neutrinos via a new light scalar mediator. We presented the benchmark parameter space which ensures that the decay lengths of the quarkonium-like bound states, hidden glueballs, and new light scalar states are shorter than the SN core size. Therefore, the stringent bound from the SN 1987A is avoided.  We have also shown that, while millicharged hidden baryons can contribute to dark matter, their abundance can account only the tiny fraction of it.

The hidden hadrons can be searched for by low-energy lepton colliders in the mono-photon plus missing energy channel. The signatures for their sub-leading decays into electrons and photons can be probed by future fixed-target experiments.

\section*{Acknowledgements}
We thank Gordan Krnjaic and Zuowei Liu for useful discussion.
The work of YB is supported by the U.S. Department of Energy under the contract DE-SC-0017647. MS and FY were supported by the Samsung Science and Technology Foundation under Project Number SSTF-BA1602-04. SL is supported by the National Research Foundation of Korea (NRF) grant funded by the Korea government (MEST) (No. NRF-2021R1A2C1005615).

\appendix

\section{Modification of $\alpha (M_Z^2)$ due to millicharged particles}
\label{app:sec:alphaMZ}
%

\begin{figure}[th!]
\centering
\includegraphics[width=0.55 \textwidth]{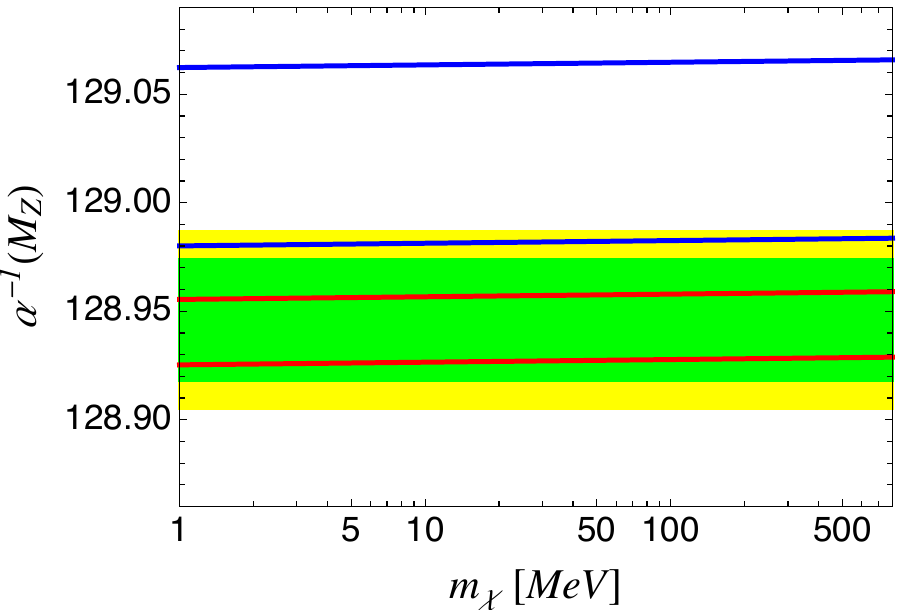}
\caption{The yellow (green) filled band represents 0.02\% (0.01\%) deviations from the central value of $\alpha^{-1}_{\text{SM}} (M_Z^2)$ obtained using $\Delta \alpha^{(5)}_{\text{had}} (M_Z^2) |_{e^+e^-} = 276.0(1.0)\times 10^{-4}$~\cite{Davier:2019can} and $N_\chi \varepsilon^2=0$.
The red band corresponds to using $\Delta \alpha^{(5)}_{\text{had}} (M_Z^2) |_{e^+e^-} = 276.0(1.0)\times 10^{-4}$ and $N_\chi \varepsilon^2 = 2.5 \times 10^{-3}$. Blue band corresponds to the case using $\Delta \alpha^{(5)}_{\text{had}} (M_Z^2) |_{\text{EW-fit}} = 270.2(3.0)\times 10^{-4}$~\cite{Crivellin:2020zul} and $N_\chi \varepsilon^2 = 2.5 \times 10^{-3}$.   
}\label{app:fig:alphaMZ}
\end{figure}

Millicharged particles contribute to both the running coupling of QED all the way up to the electroweak scale and $(g-2)_\mu$ at the same time via the common vacuum polarization. In this section we explain how millicharged particles can account for the observed anomaly in $(g-2)_\mu$ while being consistent with the precision of the QED coupling at the $Z$ mass, $\alpha(M_Z^2)$.

The fine structure constant at $M_Z$ is given by
\beqa
\begin{split}
\hspace{-5mm}\alpha^{-1} (M_Z^2) = \alpha^{-1} & \Big [ 1 - \Delta \alpha_{\text{lep}} (M_Z^2)  - \Delta \alpha^{(5)}_{\text{had}} (M_Z^2) 
 - \Delta \alpha^{\text{top}} (M_Z^2) - \Delta \alpha_\chi (M_Z^2) \Big ]~,
\end{split}
\eeqa
with $\alpha\equiv \alpha(0) = 1/137.035999084(21)$~\cite{Zyla:2020zbs}. 
Here, $\Delta \alpha_{\text{lep}} (M_Z^2)$ refers to the correction from $e,\, \mu,\, \tau$ leptons and $\Delta \alpha^{(5)}_{\text{had}} (M_Z^2)$ [$\Delta \alpha^{\text{top}} (M_Z^2)$]  the hadronic contribution from 5 (top) quark flavors. The additional new-physics contribution from mCPs, $\Delta \alpha_\chi (M_Z^2)$, at one-loop has~\footnote{For the ``heavy quark" confining hidden sector, the hidden gauge interactions provide additional sub-leading and perturbative in $\alpha_\chi$ contributions, which are ignored here.} 
\begin{equation}
 \Delta \alpha_\chi (M_Z^2) = N_\chi \varepsilon^2 \times\Delta \alpha_{\text{lep},\, l=\chi} (M_Z^2)~,
\end{equation}
with the one-loop correction from an individual lepton $l$ given by~\cite{Kuhn:1998ze}
\begin{equation}\label{app:eq:alpha:lep:oneloop}
\begin{split}
&\Delta \alpha_{\text{lep},\, l} (M_Z^2) 
= \frac{\alpha}{\pi} \left [ -\frac{5}{9} + \frac{1}{3} \ln \frac{M_Z^2}{m^2_l} - 2 \frac{m^2_l}{M_Z^2} 
+ \mathcal{O} \left ( \frac{m_l^4}{M_Z^4} \right ) \right ] ~.
\end{split}
\end{equation}
The total SM leptonic contribution at the one-loop order using Eq.~(\ref{app:eq:alpha:lep:oneloop}) is $\Delta \alpha^{\text{one-loop}}_{\text{lep}} (M_Z^2) = 314.19 \times 10^{-4}$ whereas the calculation has been performed up to four loop order with $\Delta \alpha^{\text{four-loop}}_{\text{lep}} (M_Z^2) = 314979(2) \times 10^{-7}$~\cite{Sturm:2013uka}. The contribution from top quark up to the three-loop order is $\Delta \alpha^{\text{top}} (M_Z^2) = -72.01(0.37) \times 10^{-6}$~\cite{Chetyrkin:1995ii,Chetyrkin:1996cf,Chetyrkin:1997mb,Keshavarzi:2019abf}. 

For the hadronic contribution in the SM, we take the value obtained using the dispersion relation from the cross section $\sigma(e^+e^-\rightarrow \gamma^* \rightarrow \text{hadrons})$, or $\Delta \alpha^{(5)}_{\text{had}} (M_Z^2) |_{e^+e^-} = 276.0(1.0)\times 10^{-4}$~\cite{Davier:2019can}. 
The hadronic contribution can also be obtained independently from the global fit of the electroweak precision measurements, for instance, $\Delta \alpha^{(5)}_{\text{had}} (M_Z^2) |_{\text{EW-fit}} = 270.2(3.0)\times 10^{-4}$~\cite{Crivellin:2020zul}. One notes that aforementioned two types of $\Delta \alpha_{\text{had}} (M_Z^2)$ have uncertainties of roughly 1\% which dominantly set the precision $\sim 0.01$\% of $\alpha(M_Z^2)$, and they are slightly inconsistent to each other.
In Fig.~\ref{app:fig:alphaMZ}, we show that the modification to $\alpha(M_Z^2)$ from mCPs with a mass in the range of interest and $N_\chi \cdot \varepsilon^2 = 2.5\times10^{-3}$  (red band) is below 0.01\% (the green filled band) around the central value obtained using $\Delta \alpha^{(5)}_{\text{had}} (M_Z^2)|_{e^+e^-}$ from the dispersion relation. For the purpose of the comparison, we also show the case using $\Delta \alpha^{(5)}_{\text{had}} (M_Z^2)|_{\text{EW-fit}}$ which has a slightly larger uncertainty (blue band).

For mCPs with $m_\chi = 15$ MeV and $N_\chi \varepsilon^2 = 2.5 \times 10^{-3}$ as a benchmark scenario, $\Delta \alpha_{\chi} (M_Z^2) = 122.0 \times 10^{-4} \cdot N_\chi \varepsilon^2 = 3.05 \times 10^{-5}$ which is smaller than the error size of $\Delta\alpha_{\text{had}} (M_Z^2)$. Therefore, our benchmark scenario of mCPs does not spoil the precision of $\Delta\alpha (M_Z^2)$ beyond the one set by the hadronic contribution. 

Ref.~\cite{Crivellin:2020zul} has pointed out that changing HVP to explain the muon $g-2$ excess is in conflict with the value of $\alpha(M_Z^2)$ from modern EW precision data. For the mCP explanation, we have arrived at a different conclusion. The advantage of using mCP to explain the muon $g-2$ excess while being consistent with $\alpha(M_Z^2)$ is because that one could choose a lighter mass for mCP or $m_\chi \ll m_{\pi^\pm}$ the lightest hadron mass, which can effectively enhance its contribution to the muon $g-2$. From Fig.~\ref{fig:vap:mCP:numeric}, one can observe a power-low behavior for $a_\mu$ in $m_\chi$ for $m_\chi \sim 15$~MeV. For its contribution to $\Delta\alpha (M_Z^2)$, the leading behavior is a logarithmic one [see Eq.~\eqref{app:eq:alpha:lep:oneloop}] and insensitive to $m_\chi$. 
For instance, assuming that mCPs with a mass of $\sim 100$ MeV can be treated to represent the contribution to $(g-2)_\mu$ from typical QCD light hadrons, the ratio of $\Delta a_\mu (m_\chi = 15\ \text{MeV})$ to $\Delta a_\mu (m_\chi = 100\ \text{MeV})$ is 10.8 (see Fig.~\ref{fig:vap:mCP:numeric}). This ratio gets bigger for a larger benchmark hadron mass. This indicates that the contribution to $\Delta a_\mu$ from our benchmark millicharged particle of a mass $m_\chi \sim 15$ MeV is enhanced, compared to that from QCD hadrons with higher masses than our mCPs.

\section{Lepton two-loop contribution to $(g-2)_\mu$}
\label{app:sec:twoloopQED}

The contribution to muon $g-2$ at the order of $(\alpha/\pi)^2$ from new contribution to the photon vacuum polarization of massive charged fermions (see the diagram in Fig.~\ref{fig:vap:mCP}) is given by~\cite{Elend:1966,Passera:2004bj,Passera:2006gc}
\beqa
 a_\mu^{(4)}(\text{vap},\, l) = A_{2,\, \text{vap}}^{(4)} (m_\mu/m_l) \left ( \frac{\alpha}{\pi} \right )^2~,  
\eeqa
where `vap' (superscript 4) denotes vacuum polarization (order in the coupling $e$). One has~\cite{Passera:2006gc}
\beqa\label{app:eq:QED:twoloop:full}
\begin{aligned}
  A_{2,\, \text{vap}}^{(4)} (1/x) =& - \frac{25}{36} - \frac{\ln x}{3} + x^2 ( 4+ 3 \ln x )  
  + \, x^4 \left [ \frac{\pi^2}{3} - 2 \ln x \ln \left ( \frac{1}{x} - x \right ) 
  - \text{Li}_2(x^2) \right ]
  \\
  & + \frac{x}{2} (1-5x^2) \left [ \frac{\pi^2}{2} - \ln x \ln \left ( \frac{1-x}{1+x} \right ) - \text{Li}_2 (x) +\text{Li}_2 (-x) \right ] \,,
\end{aligned}
\eeqa
where $x = m_l/m_\mu$ ($m_l$ as the mass of the virtual lepton in the loop) and $\text{Li}_2 (z) = - \int_0^z (dt/t) \ln (1-t)$.

\section{Hidden confining gauge sector}
\label{app:sec:darkQCD}
We assume that a millicharged particle $\chi$ transforms as the fundamental representation under the hidden confining gauge group $SU(N_\chi)$ with the confinement scale $\Lambda_\chi \sim 1$~MeV. The running gauge coupling $\alpha_\chi(\mu) \equiv g_\chi^2(\mu)/4\pi$ of the hidden gauge group $SU(N_\chi)$ is roughly given by $\alpha_\chi (\mu) \approx 6\pi/(11 N_\chi)[1/\ln(\mu/\Lambda_\chi)]$ by neglecting the threshold effect of $m_\chi$ and assuming a large $N_\chi$. The combination $N_\chi\,\alpha_\chi (\mu)$ can be taken to be order-one at $\mu \sim m_\chi > \Lambda_\chi$.

The $\chi$-Upsilon meson $\Upsilon_\chi$ as a $\chi\bar{\chi}$ bound state can be formed from the binding force mediated by hidden gluons. This is analogous to the $\Upsilon$ state in the SM QCD as the $b\bar{b}$ bound state where the bottom quark mass $m_b$ is bigger than the QCD confinement scale. The $\Upsilon_{\chi}$ hadron can decay into the electron pair via a virtual  photon with the decay width given by
\begin{equation}\label{eq:upsilonDwidth}
\Gamma ( \Upsilon_\chi \rightarrow e^+e^-) = N_\chi \frac{16 \pi \alpha^2 \varepsilon^2}{3} \frac{|\psi_{\Upsilon_\chi}(0)|^2}{M^2_{\Upsilon_\chi}}~,
\end{equation}
where the $\Upsilon_\chi$ mass can be taken to be roughly $m_{\Upsilon_\chi} \sim 2 m_\chi$ for a weakly bounded state. The wave function at the origin is approximated by $|\psi_{\Upsilon_\chi}(0)|^2 \approx [C_F \alpha_\chi(\mu_{\rm Bohr})\, \mu_\chi]^3$ where $\mu_\chi = m_\chi/2$ is the reduced mass of the $\chi\bar{\chi}$ pair and $\mu_{\rm Bohr}$ is the Bohr momentum, by assuming the Coulomb-like potential (ignoring the linear potential term), $V_\chi(r) \approx -C_F \alpha_\chi/r$. Here, $C_F = T_F (N_\chi^2-1)/N_\chi$ and $\text{tr} (t_a t_b) = T_F\, \delta_{ab} =\frac{1}{2}\, \delta_{ab}$.

Similarly to the QCD $\Upsilon$, $\Upsilon_\chi$ has a larger decay rate into hidden gluonic final states (also due to no $\varepsilon$ suppression), which then subsequently hadronizes into hidden glueball states since there is no light hidden pion in this model. These hidden glueballs would eventually decay to SM photons or electrons with order-one branching fractions through $\chi$ loop unless there exist other decay channels. Therefore, $\Upsilon_\chi$ should be subject to stringent constraints from experiments searching for dark photon. 

To also ameliorate the tension between the direct detection for $\Upsilon_\chi$ and the neutrino observation from SN1987, we design a model to have $\Upsilon_\chi$ dominantly decay into neutrinos. We introduce a real pseudo scalar field $S$ that couples to $\chi$ via a Yukawa coupling and to neutrinos via an $L$ (lepton number)-violating dimension-6 operator,
\begin{equation}\label{eq:SLHsq:dim6}
  \frac{S(LH)^2}{\Lambda^2}~,
\end{equation}
where the cutoff $\Lambda$ needs to be parametrically lower than the typical $L$-violating scale in the SM with a dimension-5 neutrino mass operator, or $\Lambda \ll \Lambda^{\text{SM}}_{\slashed{L}} \sim 10^{14}$ GeV to make the process of $S \rightarrow \gamma\gamma$ (mediated from a $\chi$ loop) subdominant. A common low $L$-violating scale will cause the Weinberg operator $(LH)^2$ to generate unacceptably large neutrino masses. The separation between two types of $L$-violating scales can be achieved by introducing an (approximately) discrete symmetry, under which fields transform
\begin{equation}\label{eq:dsym:breaking}
\begin{split}
   S \rightarrow -S~,\quad
   \chi_{L/R} \rightarrow \chi_{R/L}~, 
   L \rightarrow i L~,\quad 
   e_R \rightarrow i e_R~,\quad
   H \rightarrow H~.
\end{split}
\end{equation}
Under this discrete transformation in Eq.~(\ref{eq:dsym:breaking}), the following Lagrangian terms
\begin{equation}
  - i y_s S \bar{\chi}\gamma_5 \chi  - m_\chi \bar{\chi} \chi
  - ( y_l \bar{L} H e_R + \text{h.c.}) + \frac{S(LH)^2}{\Lambda^2}~,
\end{equation}
are invariant whereas the Weinberg neutrino-mass operator is not. 

The process of $S\rightarrow \gamma\gamma$ is subdominant compared to the process of $S \rightarrow \nu\nu$ by requiring
\begin{equation}
\begin{split}
 &\Gamma(S \rightarrow \nu\nu) \approx \frac{1}{16\pi} \left ( \frac{v^2}{2 \Lambda^2} \right )^2 m_S
 \gg \ \Gamma(S \rightarrow \gamma\gamma)  =  \frac{\alpha^2}{256 \pi^3} \frac{y_s^2}{2} \frac{m^3_S}{m_\chi^2}  \left | N_\chi  \varepsilon^2 F (\tau) \right |^2~
 \\[3pt]
& 
\hspace{7.3cm} \rightarrow \frac{\alpha^2}{256 \pi^3} \frac{y_s^2}{2} \frac{m^3_S}{m_\chi^2}  \left | N_\chi  \varepsilon^2 \frac{4}{3} \right |^2 \quad \left ( \tau \gg 1 \right )~,
\end{split}
\end{equation}
where the loop function is given by $F(\tau) = -2\tau [1 + (1- \tau ) f(\tau)]$ with $\tau = 4 m_\chi^2/m_S^2$ and $f(\tau) = [\sin^{-1}(1/\sqrt{\tau})]^2$ (for $\tau > 1$).
It gives rise to the inequality,
\begin{equation}
  N_\chi \varepsilon^2 \ll \frac{3 \pi}{\sqrt{2}\, y_s \alpha} \frac{m_\chi}{m_S} \left ( \frac{v}{\Lambda} \right )^2~.
\end{equation}
$S$ is also required to decay within the SN core with a radius of $\sim 1$ km, or
\begin{equation}
\begin{split}
\label{eq:S-lifetime}
 &c\tau_0(S) = 1/\Gamma(S \rightarrow \nu\nu) = 
0.9 \,\mbox{cm} 
 \times \left ( \frac{1\, \text{MeV}}{m_S} \right ) \left ( \frac{\Lambda}{3\times 10^4\, \text{GeV}}
\right )^4~.
\end{split}
\end{equation}
With the above choice of $\Lambda$ and $m_S$, $S$ decays into neutrinos well before the BBN time.

While a light pseudo-scalar $S$ with the discrete symmetry in Eq.~(\ref{eq:dsym:breaking}) makes a $0^{++}$ hidden glueball ($\mathcal{G}^{0^{++}}_\chi$) lifetime short enough so that it decays within the SN core, $\Upsilon_\chi$ still has an order-one branching fraction to the $0^{-+}$ hidden glueball ($\mathcal{G}^{0^{-+}}_\chi$). This state could mainly decay into photons with a small decay width, for instance, via the dimension-8 operator $G_\chi\widetilde{G}_\chi F\widetilde{F}$ (with a similar size for $\mathcal{G}^{0^{++}}_\chi$). To make this state decay fast, we introduce a small discrete-symmetry-breaking Yukawa interaction 
\begin{equation}
  - \kappa_s y_s S \bar{\chi}\chi~,
\end{equation}
which can mediate $\mathcal{G}_\chi^{0^{-+}}$ decaying into neutrinos via intermediate $S$'s. 
For instance, a $\mathcal{G}_\chi^{0^{-+}}$ hidden glueball can decay to two $S$'s via the dimension-6 operator, 
\begin{equation}
 N_\chi \frac{g_\chi^2 \kappa_sy_s^2}{16\pi^2 m_\chi^2} S^2 G_{\chi\, a\,\mu\nu} \widetilde{G}_\chi^{a\,\mu\nu}~.
\end{equation}
The dominant decay channel has
\beqa
\Gamma(\mathcal{G}^{0^{-+}}_\chi \rightarrow SS) \simeq  \frac{1}{8\pi} \left ( \frac{N_\chi\alpha_\chi\,\kappa_s\, y_s^2}{4\pi m_\chi^2} \right )^2 \Lambda_\chi^5 ~,
\eeqa
with non-perturbative order-one numbers neglected here. The sub-leading visible channel has a decay width of 
\beqa
\Gamma(\mathcal{G}^{0^{-+}}_\chi \rightarrow \gamma\gamma) \simeq  \frac{1}{8\pi} \left ( \frac{N_\chi\alpha_\chi\, \alpha \varepsilon^2}{m_\chi^4} \right )^2 \Lambda_\chi^9 ~.
\eeqa
As long as the inequality,
\begin{equation}
   \frac{\varepsilon}{\sqrt{\kappa_s}} \ll \frac{y_s}{\sqrt{4\pi \alpha}}\frac{m_\chi}{\Lambda_\chi}~,
\end{equation}
is satisfied (trivially satisfied for the parameter space of interest), the decay width of $\mathcal{G}_\chi^{0^{-+}} \rightarrow SS$ dominates over  $\mathcal{G}_\chi^{0^{-+}} \rightarrow \gamma\gamma$. In this limit, the visible branching ratio of $\mathcal{G}_\chi^{0^{-+}}$ is 
\beqa
\label{eq:0-+_branching}
\mbox{BR}(\mathcal{G}^{0^{-+}}_\chi \rightarrow \gamma\gamma) \simeq \frac{(4\pi)^2\varepsilon^4\,\alpha^2\Lambda_\chi^4}{\kappa_s^2\,y_s^4\,m_\chi^4} ~. 
\eeqa
The decay length of $\mathcal{G}_\chi^{0^{-+}}$ is given by
\begin{equation}
\begin{split}
\label{eq:0-+lifetime}
 &c\tau_0(\mathcal{G}_\chi^{0^{-+}}) = 1/\Gamma(\mathcal{G}^{0^{-+}}_\chi \rightarrow SS)
\\[5pt]
 & \hspace{0.3cm}  \simeq 0.9\,\mbox{cm}\times\,\left ( \frac{10}{N_\chi} \right )^2 \left ( \frac{10^{-3}}{\kappa_s} \right )^2\,  \left ( \frac{2}{y_s} \right )^4\,\left ( \frac{m_\chi}{15\, \text{MeV}}
\right )^4\, \left ( \frac{3\,\text{MeV}}{\Lambda_\chi}
\right )^5~,
\end{split}
\end{equation}
which could decay within the SN core radius $\sim 1$~km. The lifetime of $\mathcal{G}_\chi^{0^{++}}$, on the other hand, is shorter because it has no additional $\kappa_s^2$ suppression factor. While we have fixed the glueball masses around the confinement scale $\Lambda_\chi$ in our estimation for both cases, they could be heavier than $\Lambda_\chi$ and make their decay length shorter.

The small discrete-symmetry-breaking coupling  can generate a tadpole term for $S$, rendering a non-vanishing vacuum expectation value for $S$ which in turn contributes to the neutrino mass with
\begin{equation}
 \Delta m_\nu \sim  \langle S \rangle \frac{v^2}{2 \Lambda^2} \approx  \frac{\kappa_s y_s}{16 \pi^2} \frac{m_\chi^3}{m_S^2}  \frac{v^2}{2 \Lambda^2} < 0.1\ \text{eV}~.
\end{equation}
One could use the above equation to impose a theoretical constraint on the model parameter space if one requires no tuning to explain the light neutrino mass. 

For glueballs with other quantum numbers and higher masses (see \cite{Mathieu:2008me} for theoretical calculations of glueball spectra), one can also introduce additional interactions to make them decay fast enough with a decay length smaller than $\sim 1$~km. On the other hand, they have smaller branching ratios from $\Upsilon_\chi$ decays because of their heavier masses and are less harmful. 

For the $\Upsilon_\chi$ state, it mainly decays into multiple glueball states with small branching ratios into visible particles. To estimate its hidden gluonic decay width, we  adopt the formula for the gluonic decay width of the $\Upsilon$-like mesons in QCD in~\cite{Mackenzie:1981sf,Novikov:1977dq}:
\begin{equation}\label{eq:upsilon:to:hgluons}
\begin{split}
  &\Gamma(\Upsilon_\chi \rightarrow G_\chi G_\chi G_\chi ) = \frac{(N_\chi^2-4)(N_\chi^2-1)}{16 N_\chi^2}
  \frac{64}{9} \left ( \pi^2 - 9 \right ) \alpha_\chi^3(M_{\Upsilon_\chi}) \frac{|\psi_{\Upsilon_\chi}(0)|^2}{M^2_{\Upsilon_\chi}}~,
\end{split}
\end{equation}
with $|\psi_{\Upsilon_\chi}(0)|^2 \approx [C_F \alpha_\chi(\mu_{\rm Bohr})\, \mu_\chi]^3$. The $\Upsilon_\chi$ particle decays promptly. 
Similarly, the decay width to two hidden gluons and SM photon is~\cite{Mackenzie:1981sf}
\begin{equation}
\begin{split}
 & \Gamma(\Upsilon_\chi \rightarrow \gamma G_\chi G_\chi) = \frac{N_\chi^2-1}{N_\chi} 
  \frac{16}{3} \left ( \pi^2 - 9 \right ) \alpha_\chi^2(M_{\Upsilon_\chi})\, \alpha \varepsilon^2 
  \frac{|\psi_{\Upsilon_\chi}(0)|^2}{M^2_{\Upsilon_\chi}}~.
\end{split}
\end{equation}
Given three major decay channels of $\Upsilon_\chi$, two ratios, which roughly match to the branching fractions to $\gamma+X$ and $e^+e^-$, are important to check the compatibility with the constraints from dark photon detection experiments. They are
\begin{equation}
\begin{split}
&\text{BR} (\Upsilon_\chi \rightarrow \gamma G_\chi G_\chi) 
\approx \frac{12 \alpha \varepsilon^2 }{\alpha_\chi}\frac{N_\chi}{N_\chi^2-4}~,
\\[5pt]
\end{split}
\end{equation}
\begin{equation}
\label{eq:upsilon-branching-ee}
\begin{split}
&\text{BR} (\Upsilon_\chi \rightarrow e^+e^-) 
\approx \frac{12 \pi \alpha^2 \varepsilon^2}{\alpha_\chi^3} \frac{N_\chi^3}{(N_\chi^2-4)(N_\chi^2-1) (\pi^2 -9)}~,
\end{split}
\end{equation}
where $\alpha_\chi$ is evaluated at $M_{\Upsilon_\chi} \sim 2 m_\chi$. 

To recast some experimental results searching for dark photon, we also derive the kinetic mixing parameter between the $\Upsilon_\chi$ boson and photon. Following the convention 
\beqa
\mathcal{L} \supset  - \frac{1}{2} \epsilon_{\rm kin} F_{\mu\nu} {F'}^{\mu\nu} ~, 
\eeqa
with ${F'}^{\mu\nu}\equiv \partial^\mu \Upsilon_\chi^\nu - \partial^\nu \Upsilon_\chi^\mu$. By matching the decay widths of the massive gauge boson into $e^+ e^-$, one has 
\beqa
\epsilon_{\rm kin} = 2 e\varepsilon \sqrt{N_\chi} \left (\frac{N_\chi^2-1}{2N_\chi} \frac{\alpha_\chi (\mu_{\text{Bohr}}) }{4} \right )^{3/2} ~. 
\eeqa
%


\providecommand{\href}[2]{#2}\begingroup\raggedright\endgroup

\end{document}